\documentclass{mem}
\usepackage{natbib}\usepackage{txfonts}\usepackage{balance}
\usepackage{graphicx}
\usepackage[a4paper,breaklinks,dvipdfm]{hyperref}
\idline{75}{282}
\begin{document}
\def\teff{$T\rm_{eff }$}
\def\kms{$\mathrm {km s}^{-1}$}

\title{
Multifrequency Behavior of Microquasars in the GeV--TeV era: A review
}

   \subtitle{}

\author{
V. \,Bosch-Ramon\inst{1} 
          }

  \offprints{V. Bosch-Ramon}

\institute{
Dublin Institute for Advanced Studies, 31 Fitzwilliam Place, Dublin 2, Ireland
\email{valenti@cp.dias.ie}
}

\authorrunning{Bosch-Ramon}

\titlerunning{Microquasars in the GeV--TeV era}

\abstract{
Microquasars are X-ray binaries that present non-thermal radio jets. Efficient particle acceleration can take place in different regions of the jets of microquasars. The accelerated particles can
emit gamma-rays via leptonic or hadronic processes, with a complex spectral and temporal behavior. The jet termination region can be also an efficient non-thermal
emitter, as well as, in high-mass microquasars, the region of the binary system outside the jet. 
In this work, I briefly describe the physics behind the non-thermal emission observed in microquasars at different scales, focusing in the
GeV and TeV bands.
}
\maketitle{}

\section{Introduction}

Microquasars are X-ray binaries with non-thermal jets (e.g. \citealt{mir99,rib05}), being called high-mass microquasars when hosting a massive star, and low-mass microquasars
otherwise. The energy powering the non-thermal emission in microquasars can be either of accretion or black-hole rotation origin. The magnetic and kinetic power is channelled through a jet
launched from the inner regions of the accretion disk (e.g. \citealt{bla77,bla82}), and part of this power is eventually converted into relativistic particles and radiation.

For several decades, microquasars were considered strong candidates to gamma-ray sources (e.g. \citealt{cha85}; see also \citealt{cha89,lev96,par00}), but they have not become fully
recognized as powerful gamma-ray emitters until recent years, after the most recent generation of ground-based Cherenkov (HESS, MAGIC, VERITAS) and satellite-borne instruments ({\it Fermi},
{\it AGILE}) arrived. The most relevant cases are the microquasars Cygnus~X-1\footnote{This source has been detected in GeV and TeV energies with significances close, but slightly below,
5~$\sigma$, and thus these detections are still to be firmly established.} and Cygnus~X-3 \citep{alb07,sab10,tav09,abd09a,sab11}. Other sources, like for instance SS~433, Scorpius~X-1 or
GRS~1915$-$105, have been also observed in GeV and TeV energies but only upper limits have been obtained \citep{sai09,ace09,ale10,bor10}. It is noteworthy that there are four other binary
systems that may be also microquasars: LS~I+61+303 (e.g. \citealt{alb06,abd09b,pit09}), LS~5039 (e.g. \citealt{aha05,abd09c,pit09}), HESS~J0632$+$057 (e.g. \citealt{hin09,fal11,mol11}), and
1FGL~J1018.6$-$5856 (e.g. \citealt{cor11}), although they could as well host a non-accreting pulsar. Regarding the microquasar and pulsar scenarios, LS~I+61+303 and LS~5039 have been
extensively discussed in the literature (see, e.g., \citealt{bos09} and references therein).

Although the detected radio emission is already evidence of particle acceleration in microquasar jets, the finding of microquasar gamma-ray emission proves that these sources can very
efficiently channel accretion or black-hole rotational energy into radiation. In addition, together with this high efficiency, the temporal characteristics of the detected radiation may
favor leptonic models, although hadronic mechanisms cannot be discarded. Also, the extreme conditions under which gamma-rays are produced can put restrictions in the emitter structure.
Morphological studies can be also of help, since non-thermal processes can take place not only at the binary scales, but also far away (e.g. the jet termination region). Although the
complexity of microquasar phenomenology can make the characterization of the ongoing processes difficult, high quality data together with semi-analytical modeling can provide sensible
information on the non-thermal physics of the sources. Numerical calculations are also important, since they can inform about the conditions of the background plasma in which emission takes
place.

\begin{figure*} 
\center
\includegraphics[width=6cm]{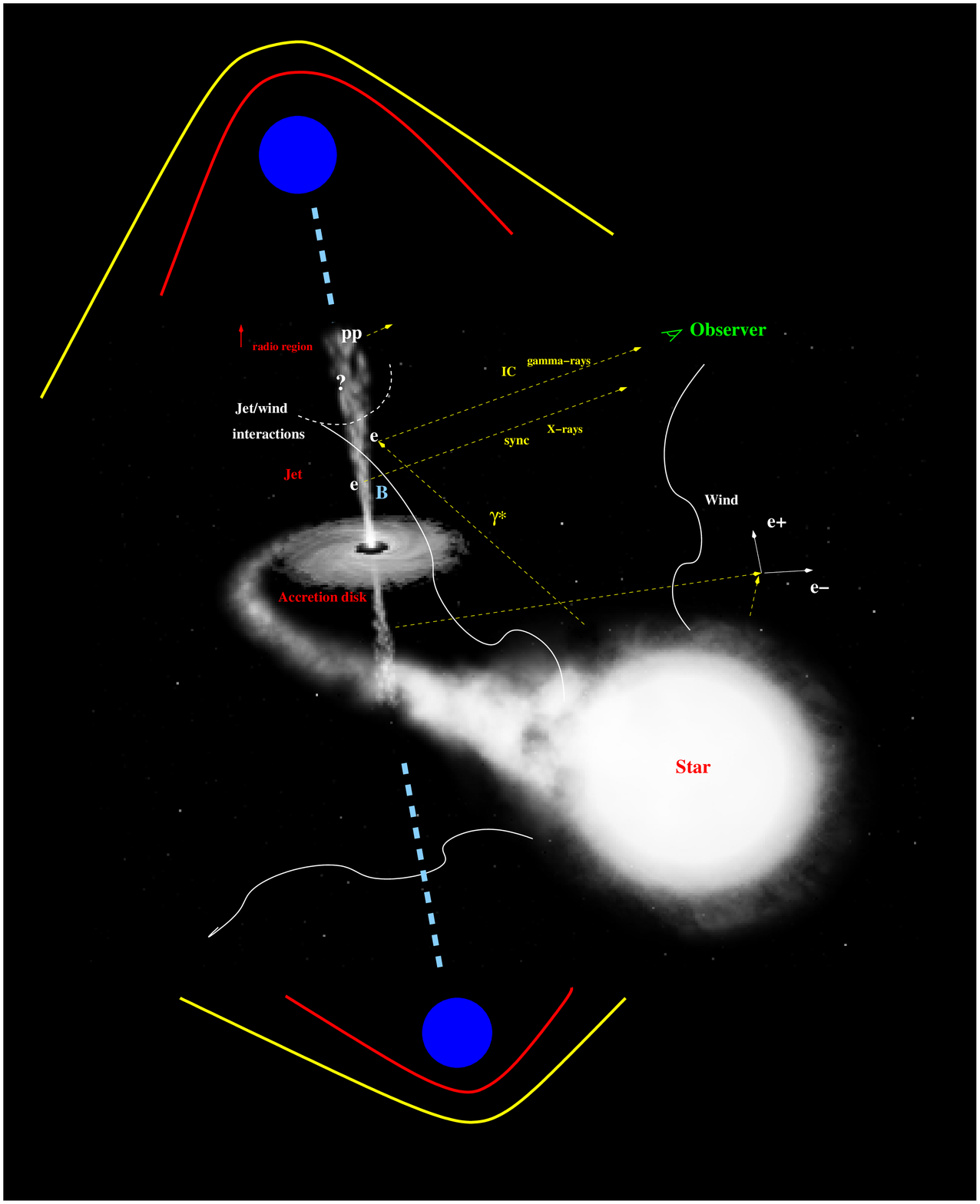}
\caption{Illustrative picture of the microquasar scenario (not to scale), 
in which relevant elements and dynamical and radiative processes in different regions are shown (background image from ESA, NASA, and F\'elix Mirabel).} 
\label{mq}
\end{figure*}

In this paper, we briefly review relevant aspects of the non-thermal emission in microquasars. We will focus mainly in the GeV and TeV energy bands, for which photon production requires
extreme conditions in these sources. In Figure~\ref{mq}, a sketch of the microquasar scenario is presented.

\section{Non-thermal emission in microquasars}

Microquasar jets can produce non-thermal populations of relativistic particles via diffusive shock acceleration or other mechanisms at different spatial scales (e.g. \citealt{rie07}). These
particles, electrons, protons or even heavy nuclei, can interact with the background matter, radiation and magnetic fields to produce non-thermal emission from radio to gamma-rays. In the
GeV and TeV bands, the most efficient process is inverse Compton (IC), but in systems with very high density regions, like SS~433, Cygnus~X-3 and possibly Cygnus~X-1, proton-proton
interactions may be also relevant. Also, in systems with very dense fields of energetic target photons, photomeson production and even photodisintegration of nuclei may be efficient. The
gamma-ray emission can take place at different scales, although certain regions suffer from strong absorption via pair creation (e.g. deep inside the system or at the jet base), and some
others may (or may not) lack enough targets (e.g. the jet largest scales). Below, we discuss farther the non-thermal phenomena at different scales in high- and low-mass microquasars. For a
general review on the efficiency of leptonic and hadronic processes under typical microquasar conditions, see \cite{bos09} and references therein.

\subsection{Microquasar emitting sites}

Different emitting regions can be considered when understanding the non-thermal emission from microquasars. Jets are the best acceleration sites given the large amount of energy that they
transport. Different forms of dissipation can take place in them through shocks, velocity gradients and turbulence (e.g. \citealt{rie07}), as well as magnetic reconnection (e.g.
\citealt{zen01}), which can lead to generate non-thermal particle populations. 

\begin{figure} 
\center
\includegraphics[angle=270,width=6.5cm]{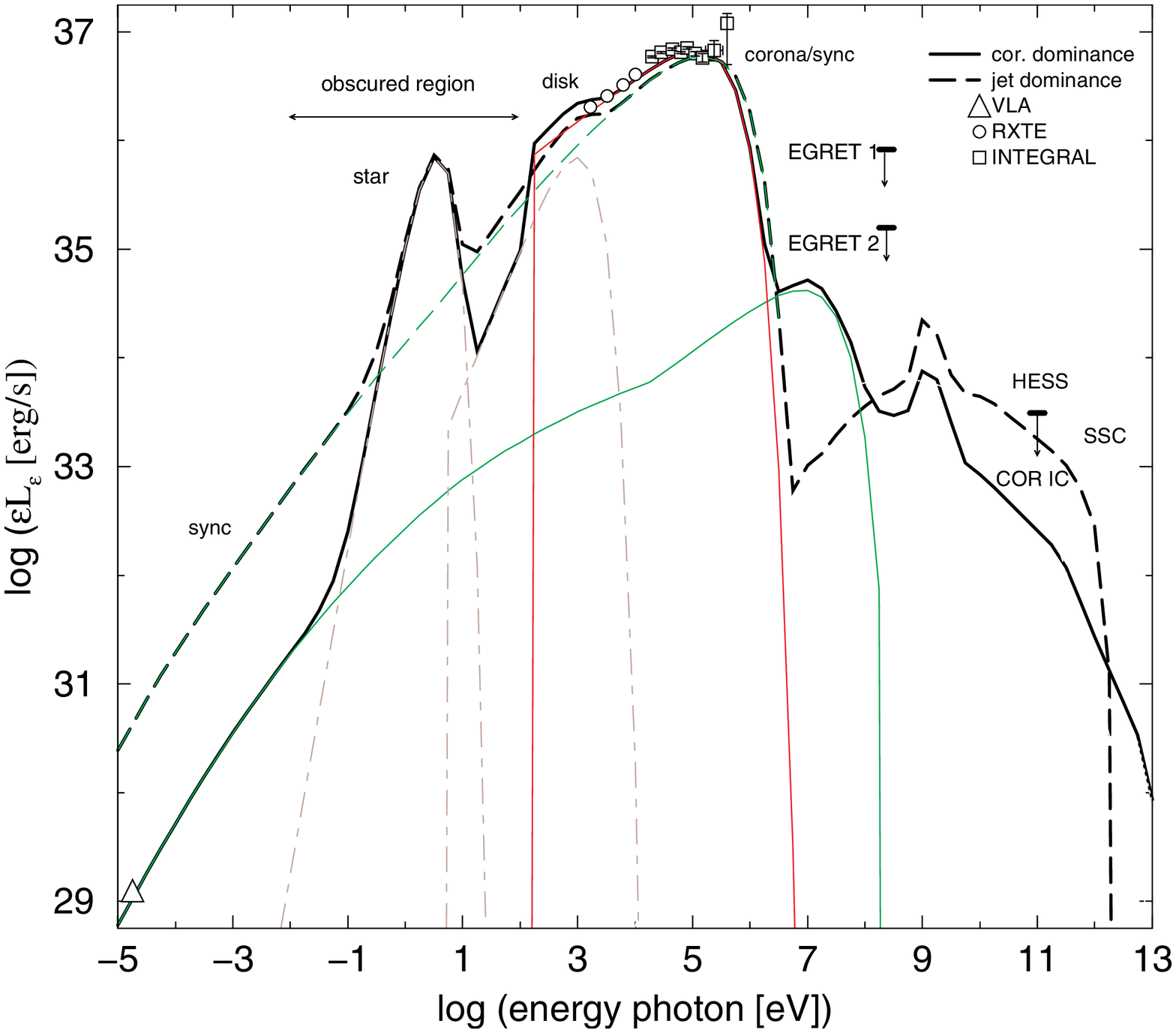}
\caption{
Computed spectral energy distribution of the non-thermal emission from 1E~1740.7$-$2942 for two situations. In one case, 
the hard X-rays come from a corona, whereas in the other, they are of synchrotron origin and come from the
jet. Gamma-ray absorption in the accretion disk and corona photon fields has been taken into account (see the two dips below and above
$\sim$ GeV energies). For details, see \cite{bos06b}.
} 
\label{lmqs}
\end{figure}

The jet formation itself, interaction with an accretion disc wind, or recollimation and internal shocks can accelerate particles at the jet base. The presence of non-thermal electrons in
the region can lead to the production of gamma rays through IC with accretion photons, or with photons produced by the same electrons via synchrotron emission (e.g. \citealt{bos06b}). The
base of the jet is possibly the region in which hadronic processes may be the most efficient, given the high density of matter and photons in there, and the hardness of the latter (e.g.
\citealt{lev01,rom08}). However, the local radiation fields could also strongly suppress the GeV emission via gamma-ray absorption and pair creation (see, e.g., \citealt{rom08,cer11}). For
low ambient magnetic fields, electromagnetic cascades can increase the effective transparency of the source to gamma-rays \citep{akh85}. An example of a (leptonic) low-mass microquasar
spectral energy distribution, with its high-energy radiation mainly coming from the base of the jet, is shown in Fig.~\ref{lmqs}.

\begin{figure*} 
\center
\includegraphics[angle=270,width=6.5cm]{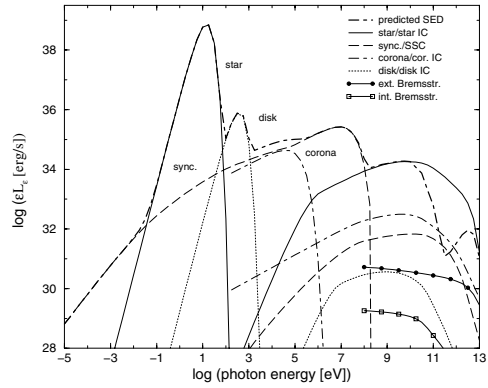}
\caption{
Computed spectral energy distribution, from radio to very high energies, for a high-mass microquasar (see \cite{bos06a}).} 
\label{hm}
\end{figure*}

In high-mass microquasars, the strong radiation and mass-loss from the star can render significant non-thermal radiation, in particular at high energies, whereas radio may be at least
partially free-free absorbed. The considered most efficient high-energy channel is typically IC with stellar photons (e.g. \citealt{bos06a}), interaction that is anisotropic and has
specific lightcurve and spectral features (e.g. \citealt{kha08}). Anisotropic IC may be behind the orbital modulation of the GeV lightcurve of Cygnus~X-3 seen by {\it Fermi}
\citep{abd09a,dub10}. Absorption of TeV emission in the stellar photon field is likely to be significant for compact high-mass systems, like Cygnus~X-3 and Cygnus~X-1. That may be the
reason why the former has not been detected in the TeV range (see \citealt{ale10} and references therein), and why the evidence of detection of Cygnus~X-1 by MAGIC may imply an emitter
outside the binary system (see \citealt{bos08a}). As before, for low enough magnetic fields (see \citealt{kha08}), electromagnetic cascades can increase the effective transparency of these
two sources (see, e.g., \citealt{bed07,ore07,bed10}). As discussed below, the role of pair creation cannot be neglected in the context of broadband non-thermal emission. At the binary
scales, absorption of GeV photons is not expected since this band is below the gamma-ray energy threshold for pair creation, around $\sim 10-100$~GeV for stellar photons peaking in the UV.
Proton-proton, photomeson production and photodisintegration have also been proposed as possible mechanisms of gamma-ray emission at these scales (e.g. \citealt{rom03,aha06,bed05}). An
example of a (leptonic) spectral energy distribution of a high-mass microquasar is shown in Fig.~\ref{hm}.

\begin{figure*} 
\center
\includegraphics[width=11cm]{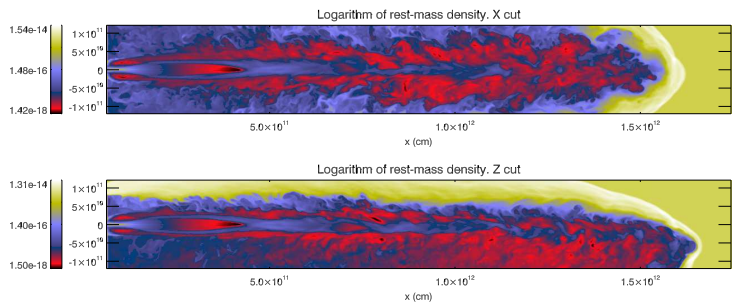}
\caption{
Density map for a high-mass microquasar jet interacting with the stellar wind, which is coming from the top of the 
image (see \cite{per10}).
} 
\label{jetint}
\end{figure*}

The interaction of the jets with the stellar wind cannot be neglected in microquasars with a massive companion. The impact of the wind on the jet triggers strong shocks, good candidates for
particle acceleration, jet bending, and potentially jet disruption (e.g. \citealt{per08,per10}). This interaction can generate high-energy emission \citep{per08}, but the specific
properties can depend on the level of inhomogeneity of the stellar wind (e.g. \citealt{ara09,ara11}; also Perucho \& Bosch-Ramon, in preparation). Figure~\ref{jetint} shows the density map
resulting from a 3-dimensional simulation of a microquasar jet interacting with the wind of the companion.

Far from the binary system, say at milliarcsecond to second scales, the jet propagates unaffected by significant external disturbances. However, there are different mechanisms that
may lead to energy dissipation, particle heating/acceleration and subsequent radiation, like velocity gradients and Kelvin-Helmholzt instabilities in the jet walls. Shear acceleration has
been proposed for instance to explain extended emission from large scale jets in microquasars and AGNs \citep{rie07}. All this could generate fresh relativistic particles that could emit in
radio by synchrotron. Very powerful ejections could also be bright enough to be detectable, from radio to gamma-rays, far away from the binary (e.g. \citealt{ato99}). 

\begin{figure} 
\center
\includegraphics[angle=0,width=6.5cm]{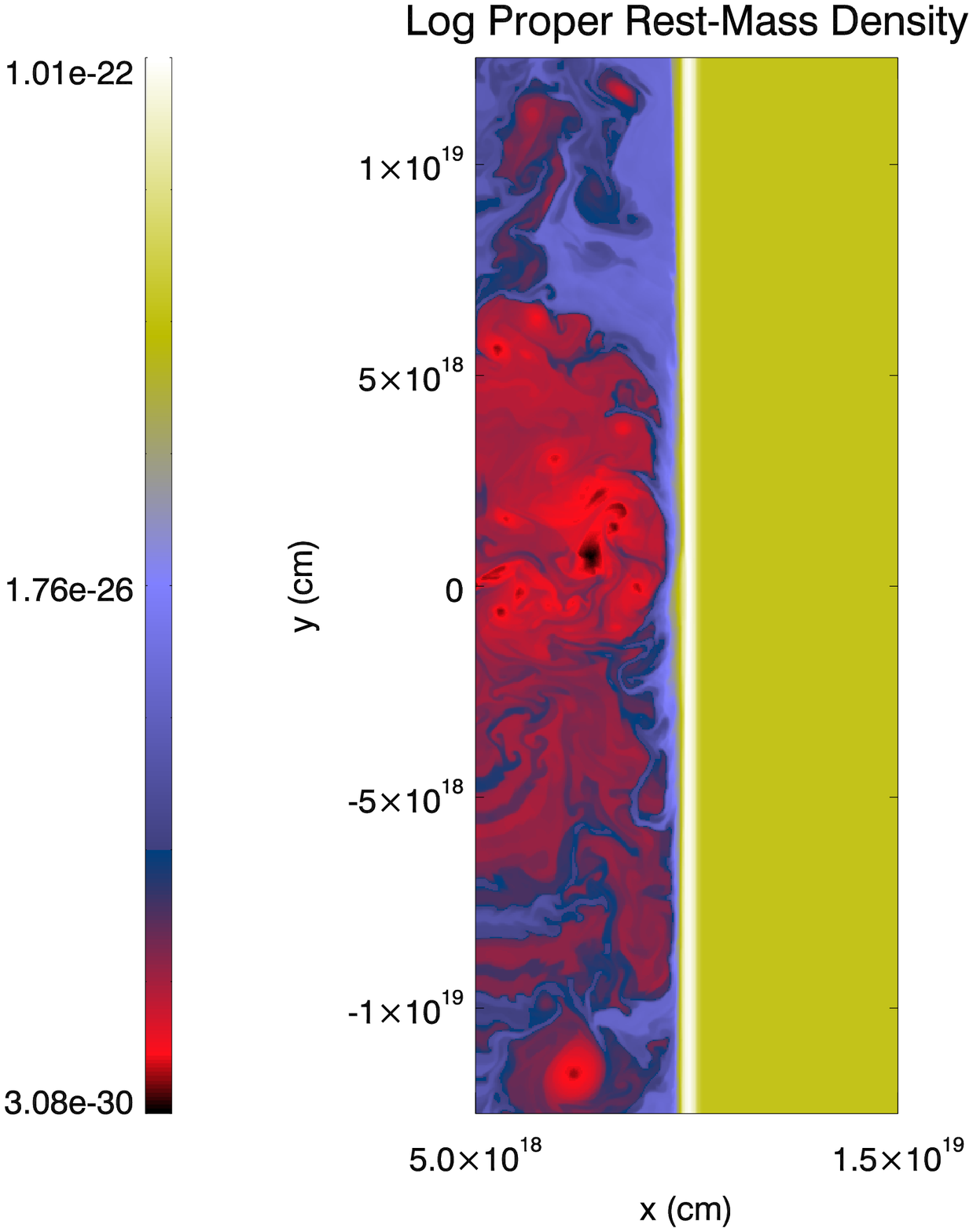}
\caption{Density map resulting from a 2-dimensional slab simulation, in which
a jet propagates in an environment characterized by the microquasar motion in the ISM. The shocked stellar wind, deflected by the microquasar 
proper motion, comes from the top and impact the jet from a side (see \citealt{bos11b}).} 
\label{ls}
\end{figure}

It is noteworthy that, unless there is not significant previous jet activity, the wall of a continuous jet, or a transient ejection, are always to encounter diluted and hot jet material.
This material was reprocessed in the jet reverse shock, where jet and ISM pressures balance, and was swept backwards filling the so-called cocoon. Only the presence of a strong wind,
either from the accretion disc or the star, can clean this material out up to a certain distance from the microquasar. However, the jet material will unavoidably end up embedded in the
cocoon plasma before the reverse shock is reached. The pressure of the cocoon can trigger a recollimation shock in the jet, which becomes collimated and suffers pinching. The jet fed cocoon
drives a slow forward shock in the ISM, much denser and cooler than the jet. This complex dynamical behavior has associated the production of non-thermal emission, which possibly may reach
gamma-ray energies. An interesting situation arises when the microquasar has a high-mass companion and the proper motion velocity is $\gtrsim 10^7$~cm~s$^{-1}$, in which case the jet can be
completely disrupted before reaching the ISM, as illustrated in Fig.~\ref{ls}. Farther discussion of jets interacting with the ISM can be found in \cite{bor09}, \cite{bos11b}, and
references therein.

\begin{figure*}
\centering
\includegraphics[angle=0, width=0.35\textwidth]{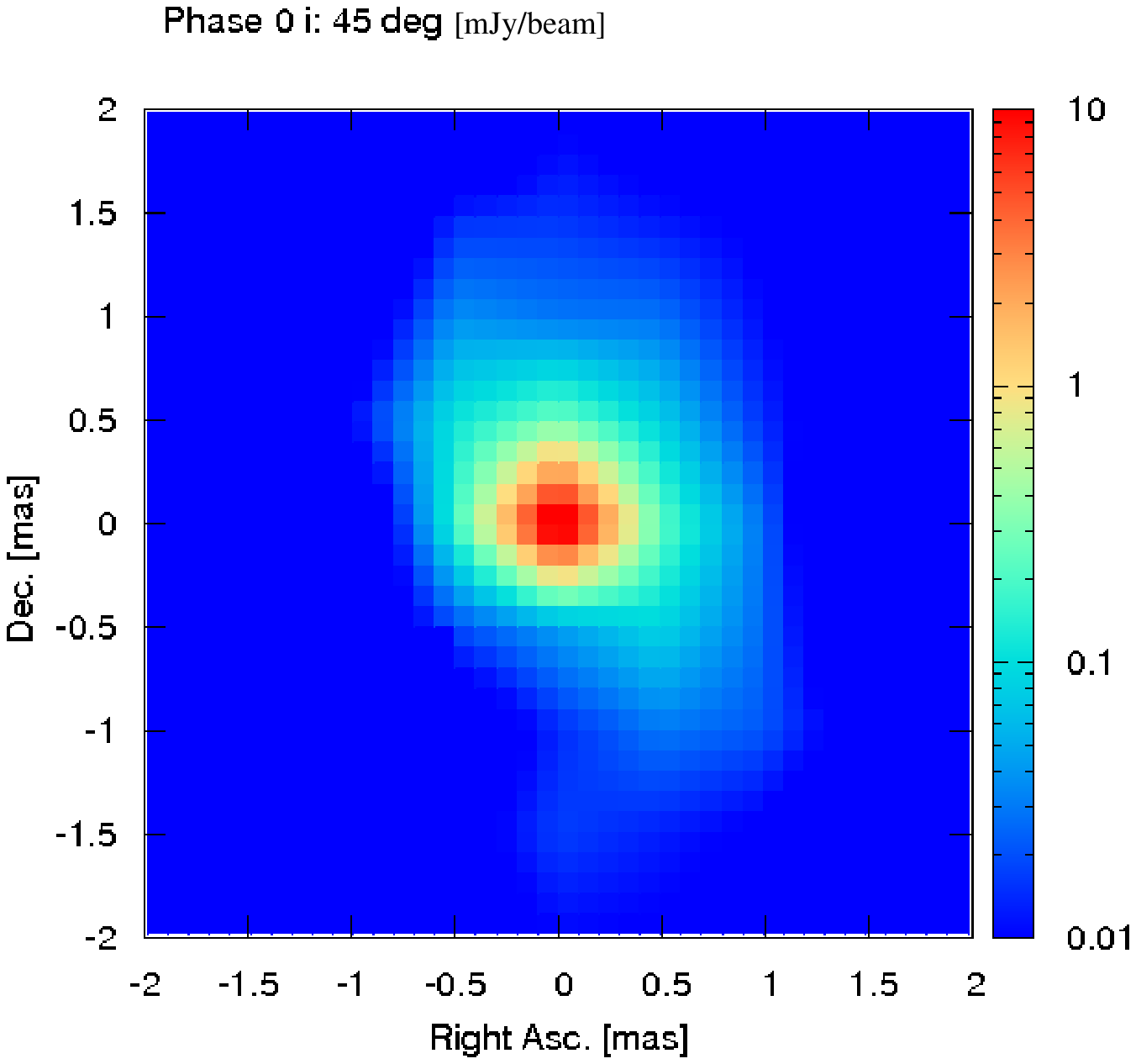}\qquad
\includegraphics[angle=0, width=0.35\textwidth]{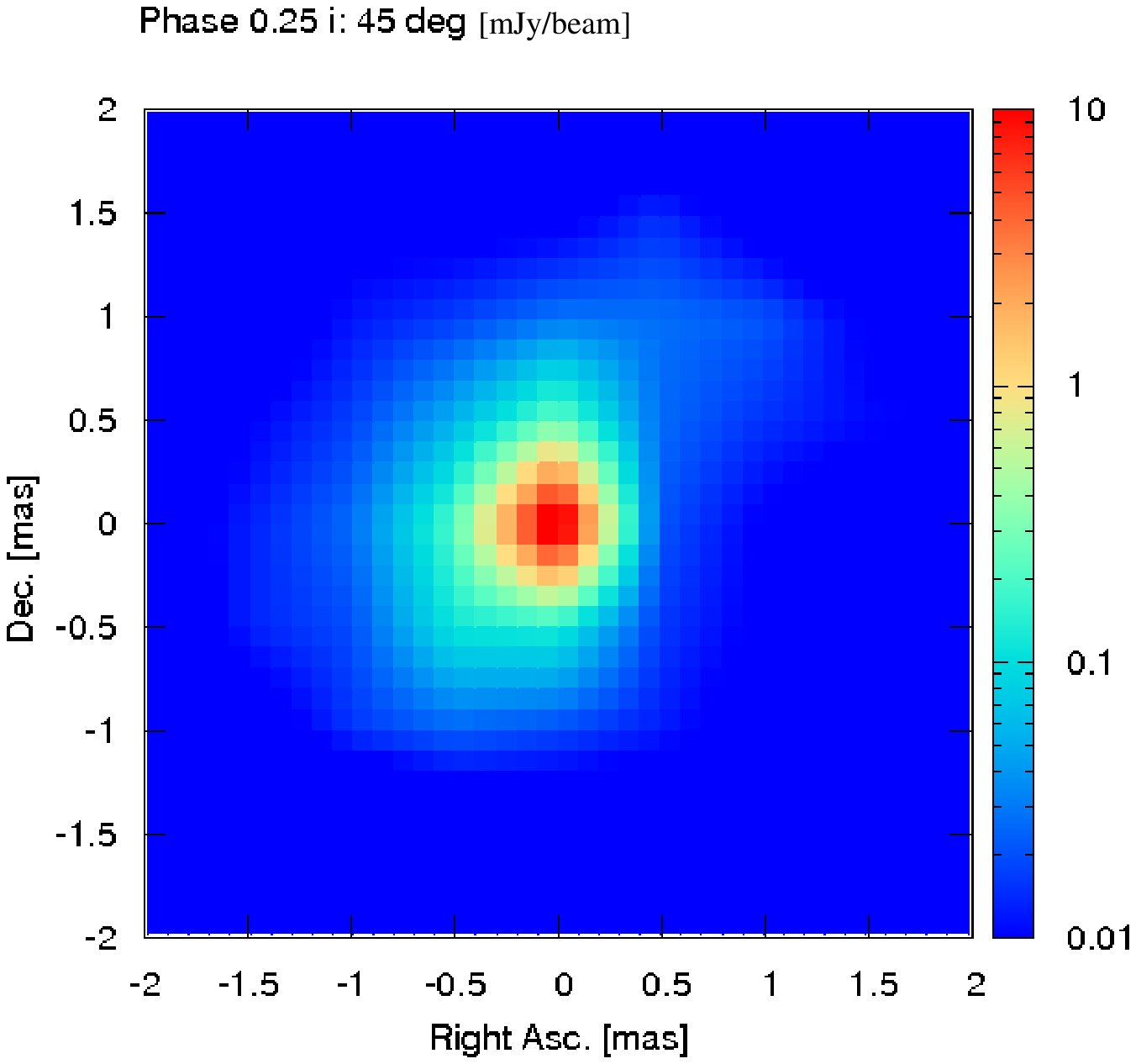}\\[10pt]
\includegraphics[angle=0, width=0.35\textwidth]{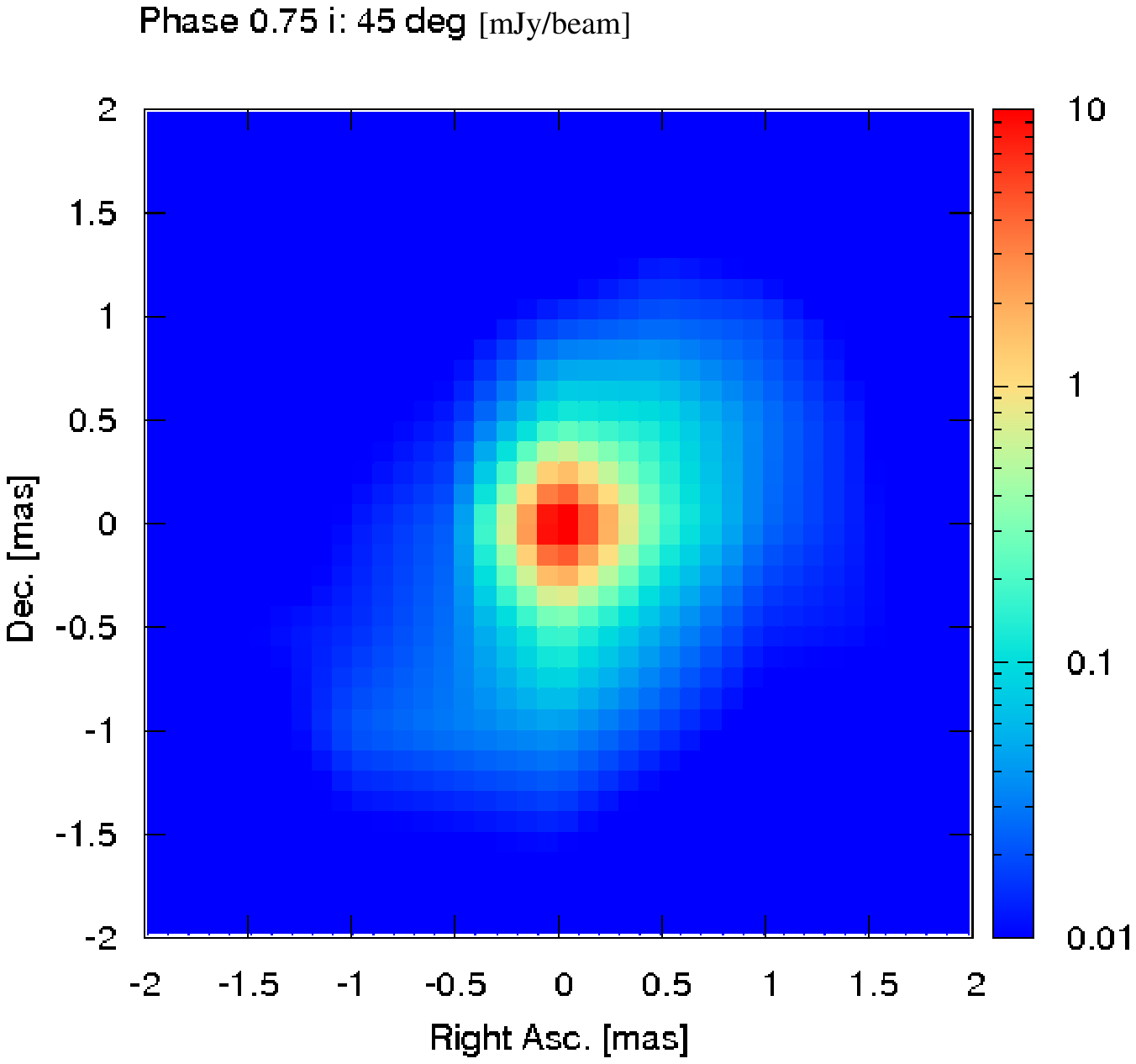}\qquad
\includegraphics[angle=0, width=0.35\textwidth]{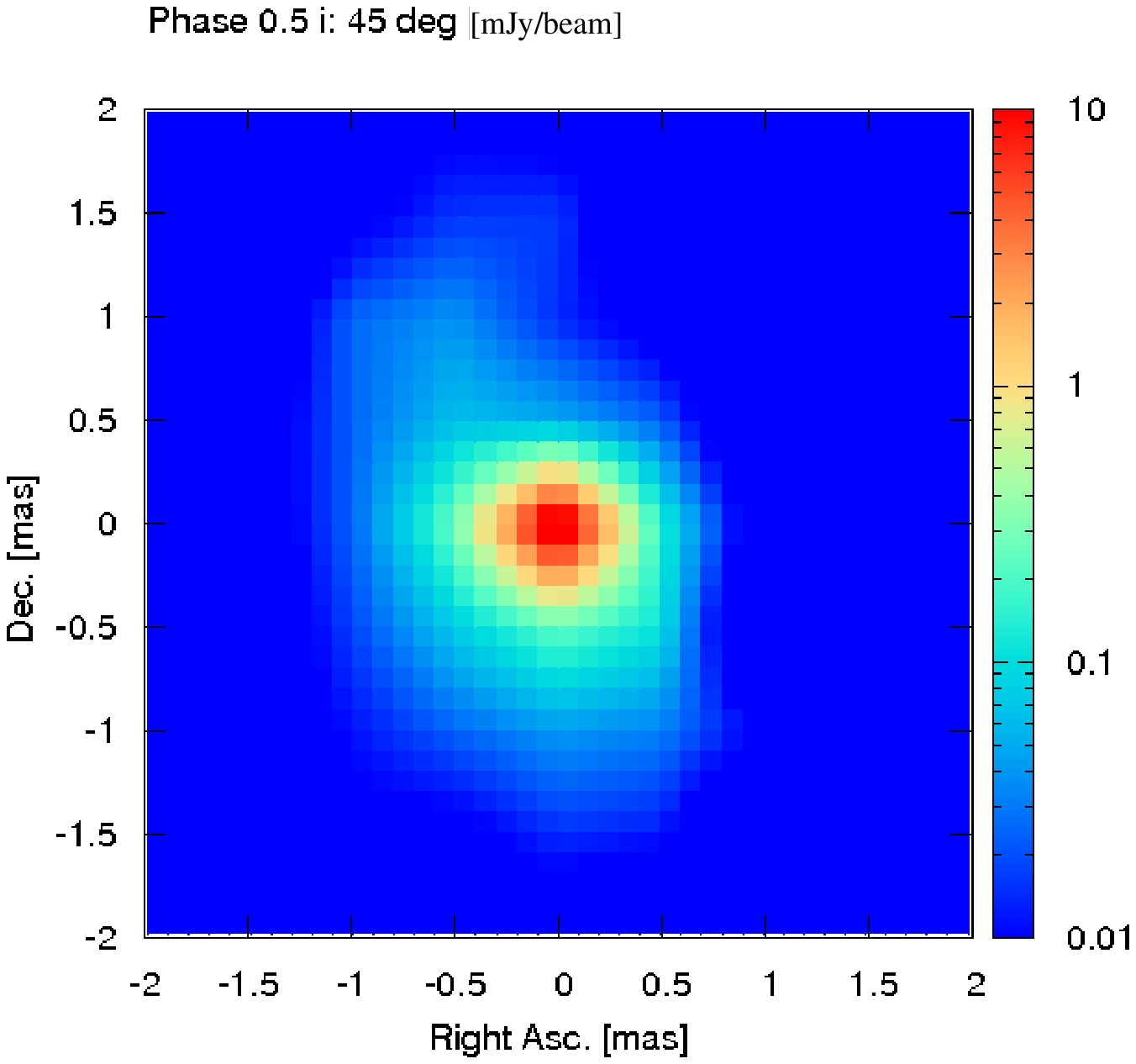}\\
\caption{Computed image, in the direction to the observer, of the 5~GHz radio emission from secondary pairs in a TeV binary for different orbital phases.
Units are given in mJy per beam, being the beam size $\sim 1$~milliarcsecond (see \cite{bos11a}).}
\label{raw}
\end{figure*}

Jets, or their termination region, are not the only possible emitting sites in microquasars. The inner regions of the accretion structures (e.g. disc, corona/ADAF and the like) may also
contain non-thermal populations of particles (e.g. \citealt{bis76,pin82,spr88,gie99,rom10}). At the binary system scales, and in particular with high-mass companions, the stellar wind is
dense, carries magnetic field, and is embedded in a dense photon bath by the star. Therefore, for those systems with very efficient particle acceleration in the jet, electrons and protons
could diffuse out of it and radiate their energy in the environment. Also, gamma-ray absorption due to pair creation in the stellar photon field can inject electrons and positrons in the
wind, also leading to broadband non-thermal emission, as shown for instance in \cite{bos08b}. This emission may be actually behind a substantial fraction of the milliarcsecond radiation in
a TeV emitting microquasar \citep{bos11a}. An example of this is shown in Fig.~\ref{raw}, in which 5~GHz maps are presented for different orbital phases in a TeV emitting binary.

\section{Final remarks}

Microquasar can efficiently accelerate particles up to very high energies and produce gamma-rays within and outside the binary system. It is however unclear currently why some sources emit
gamma-rays and others do not. A key point may be the presence of a massive star, which as discussed here can affect the jet significantly, with the formation of particle acceleration sites,
and also offering dense target photon and matter fields, suitable for gamma-ray production at the binary scales. Low-mass microquasars could in principle produce gamma-rays, and a reason
for remaining undetected yet may be that their GeV emission is too dim for the present instrumentation. At TeV energies, in the context of leptonic models, strong synchrotron cooling and IC
scattering deep in the Klein-Nishina regime (as expected in the hard accretion photon fields of low-mass microquasars) may also prevent their detection in TeV. The lack of a dense stellar
wind in low-mass systems makes also a difference with high-mass ones. It is worth noting that the four mentioned gamma-ray microquasar candidates harbor massive stars, which seems to be a
common feature in most of the known gamma-ray binaries (not only microquasars). At the jet base, the situation seems to be basically the same in high- and low-mass systems. In both object
types, the compactness of the region could imply that gamma rays are absorbed. Presently, gamma rays have not been detected at the largest scales, and thus it is not clear whether there is
an intrinsic difference between high- and low-mass microquasars at these scales, which may be the case accounting for their different environments.

\begin{acknowledgements}
I want to thank the organizers for their kind invitation.
The research leading to these results has received funding from the European
Union
Seventh Framework Program (FP7/2007-2013) under grant agreement
PIEF-GA-2009-252463.
V.B.-R. acknowledges support by the Spanish Ministerio de Ciencia e 
Innovación (MICINN) under grants AYA2010-21782-C03-01 and 
FPA2010-22056-C06-02.
\end{acknowledgements}

\bibliographystyle{aa}

\bigskip
\bigskip
\noindent {\bf DISCUSSION}

\bigskip
\noindent {\bf WOLFGANG KUNDT's Comment:} In your thoughtful review on VHE radiation from microquasars you also addressed the expected VHE emission region. How realistic was your cartoon? In my understanding of all the jet sources, their high-energy leptons are created by magnetic reconnection in the distorted magnetosphere of its central rotator, and post-accelerated by its outgoing frequency waves, on scales $\la$ that of the Blandford \& Rees de-Laval nozzle, some $10^{15\pm 1}$~cm. This scale was also found by Martin Kluczykont for M~87. It should exceed that of the accretion disc.

\bigskip
\noindent {\bf VALENT\'I BOSCH-RAMON:} In standard models of jet formation, the jet launching region is $\sim 10^2-10^3$~$R_{\rm Sch}$, so $\sim 10^9$~cm for a stellar mass black hole. In
that framework, within that region the jet would be magnetically dominated and magnetic reconnection may be important, but farther, particle 
acceleration is likely to take place through kinetic energy dissipation, via some diffusive acceleration process of the Fermi type. Of course, the issue is still open.

\bigskip
\noindent {\bf IMMACOLATA DONNARUMMA:} How could the GeV detection of Cygnus~X-1 challenge the theoretical interpretation of microquasar activity? May you compare the case of Cygnus~X-1 with the one of Cygnus~X-3?

\bigskip \noindent {\bf VALENT\'I BOSCH-RAMON:} The detection of GeV emission from Gygnus~X-1 shows that the emitter is not close from the accretion disc nor the jet base. It also shows
that the luminosity budget is very high. All this also applies to Cygnus~X-3, but in this object this GeV lightcurve is modulated along the orbit, very likely of leptinic IC origin. Such evidence for leptonic emission still lacks in the case of Cygnus~X-1. The fast variability or orbital modulation imply for both sources that the GeV emission can hardly take place far from the binary.

\end{document}